\documentclass[11pt]{article}
\usepackage{jheppub,amsmath, amsthm, amssymb,slashed,url}
\usepackage{graphicx}
\usepackage{epstopdf}

\usepackage{etoolbox}
    \makeatletter
    \patchcmd{\maketitle}{\@fpheader}{}{}{}
    \makeatother

\newcommand{\Choose}[2]{{\begin{pmatrix} {#1} \\ {#2} \end{pmatrix}}}

\def\oneht{\textstyle{1\over 2} }
\def\threeht{\textstyle{3\over 2} }
\def\onefourth{\textstyle{1\over 4} }

\def\OMIT#1{{}}

\newcommand{\beq}{\begin{equation}}
\newcommand{\eeq}{\end{equation}}
\newcommand{\beqa}{\begin{eqnarray}}
\newcommand{\eeqa}{\end{eqnarray}}

\newcommand{\nn}{\nonumber}

\begin{document}

\dedicated{NT@UW-18-09}
\title{Dimensional crossover in non-relativistic\\ effective field theory}
\vskip 0.5cm
\author{\bf Silas R.~Beane and Murtaza Jafry}
\affiliation{Department of Physics, University of Washington
Seattle, WA 98195.}

\vphantom{} \vskip 1.4cm \abstract{Isotropic scattering in various
  spatial dimensions is considered for arbitrary finite-range
  potentials using non-relativistic effective field theory. With
  periodic boundary conditions, compactifications from a box to a
  plane and to a wire, and from a plane to a wire, are considered by
  matching S-matrix elements. The problem is greatly simplified by
  regulating the ultraviolet divergences using dimensional
  regularization with minimal subtraction. General relations among
  (all) effective-range parameters in the various dimensions are
  derived, and the dependence of bound states on changing
  dimensionality are considered. Generally, it is found that
  compactification binds the two-body system, even if the
  uncompactified system is unbound.  For instance, compactification
  from a box to a plane gives rise to a bound state with binding
  momentum given by $\ln \left(\oneht\left(3+\sqrt{5} \right) \right)$
  in units of the inverse compactification length. This binding
  momentum is universal in the sense that it does not depend on the
  two-body interaction in the box. When the two-body system in the box
  is at unitarity, the S-matrices of the compactified two-body system
  on the plane and on the wire are given exactly as universal
  functions of the compactification length.}

\maketitle



\section{Introduction}

\noindent Recent experimental advances have brought remarkable control
to bear on atomic systems~\cite{RevModPhys.80.885}. Both the strength
of the inter-atomic interaction and the dimensionality of space can be
altered in ways that require an understanding of non-relativistic
quantum mechanics as the interaction potential and the dimensionality
of space are varied. Recent investigations of the phase diagram of
Bose gases as the number of spatial dimensions is continuously varied
have made use of relations among the scattering lengths in various
dimensions~\cite{PhysRevLett.81.938,2001PhRvA..64a2706P,PhysRevLett.100.170404,PhysRevA.93.063631,2018arXiv180601784I}.
In addition, there has been interest in investigating the three-body system
and the Efimov effect as spatial dimensions are varied~\cite{0953-4075-48-2-025302,0953-4075-51-6-065004}.
In cold-atom experiments, confinement of a dimension is typically
achieved using trapping potentials. However, here
toroidal confinement will be considered in a general setup,
where the fundamental assumption made is that the relevant potentials
are of finite range. This allows a formulation of the problem in terms
of an effective non-relativistic action which is an expansion in local
operators. A primary difficulty in dealing with local operators is the
renormalization which is necessary to deal with the highly singular
nature of the delta-function interactions. While the observable
physics which results from an investigation of these interactions is,
of course, not dependent on the manner in which the theory is
regulated, it is highly beneficial to choose the regularization and
renormalization scheme wisely, particularly if one is interested in a
general analysis which holds for any finite range potential. The
technology of effective field theory (EFT) is well known to be suited
to the task\footnote{For reviews, see,
  Ref.~\cite{Kaplan:2005es,Braaten:2000eh}.}.

A useful way of expressing the S-matrix for two-body scattering is via
the effective range expansion, which is valid at momenta small
compared to the inverse of the range of the interaction. Effective
range theory is therefore a natural means of expressing observables
calculated from EFT. It is a straightforward exercise to obtain the
effective range expansions to all orders in various spatial
dimensions. The question of interest here is what occurs in the
presence of boundaries which continuously interpolate between
dimensions. An elegant way of doing this is by imposing a boundary
with periodic boundary conditions and then shrinking the boundary. In
this manner one compactifies a three-dimensional box to a
two-dimensional plane or a one-dimensional wire and expresses the
one- and two-dimensional effective range parameters in terms of the
three-dimensional parameters. In similar fashion, the one-dimensional
effective range parameters can be expressed in terms of the
two-dimensional parameters. Of course, all of these relations are
dependent on the initial geometry. The main mathematical
characteristic of the toroidal compactifications is the presence of
the Riemann zeta function and related functions and several
interesting approximate relations among Riemann zeta functions of odd
integer and half-integer argument are found to emerge naturally from
the compactification scheme. An interesting theoretical scenario which
is straightforward to investigate in the general formulation occurs
when the initial system in three spatial dimensions is at
unitarity. In this case, due to the absence of a scale in the initial
configuration, the presence of a boundary gives rise to S-matrices
that are universal in the sense that they are exactly calculable in
terms of the confinement length.

This paper is organized as follows. Sec.~\ref{sec:EFT} reviews the EFT
relevant to the interactions of non-relativistic identical bosons in
$d$ spacetime dimensions. The general form of the two-body isotropic
scattering amplitude is constructed. This analysis is greatly
simplified by regulating the theory using dimensional regularization
(DR) with minimal subtraction ($\overline{MS}$). In
Sec.~\ref{sec:IsoSCatt}, the special cases with $d=2$, $d=4$, and
$d=3$ are reviewed and the effective range expansions are
defined. Special attention is given to the case $d=3$ as only this
case experiences non-trivial renormalization. Sec.~\ref{sec:compact}
contains the main results of the paper. Starting from a $d=4$ box with
periodic boundary conditions, compactification to $d=3$ and to $d=2$
is considered. All effective range parameters in $d=3$ and $d=2$ are
expressed in terms of the $d=4$ effective range parameters. The
special case of compactification when the $d=4$ theory possesses
Schr\"odinger symmetry is considered. Then, starting from a $d=3$
square with periodic boundary conditions, compactification to $d=2$ is
considered. All effective range parameters in $d=2$ are expressed in
terms of the $d=3$ effective range parameters. With the various
results in hand, a comparison is performed of the one-step versus
two-step compactification from $d=4$ to $d=2$.  Finally,
Sec.~\ref{sec:Conc} summarizes the main points of the paper.

\section{Effective field theory}
\label{sec:EFT}

\noindent This section reviews EFT technology which is helpful in 
deriving a general expression for the isotropic scattering phase shift
in any number of spatial dimensions~\cite{Beane:2010ny}.  If one is
interested in non-relativistic scattering at low energies, an
arbitrary interaction potential that is of finite range may be
replaced by an tower of contact operators\footnote{In coordinate space
  this corresponds to a sequence of delta-functions and their
  derivatives.}, whose coefficients are determined either by matching
to some known underlying theory, or by fitting to experimental
data~\cite{Braaten:1996rq,Kaplan:1998tg,vanKolck:1998bw}. The crucial
point is that at low energies only a few of the contact operators will
be important. The EFT of bosons, interacting isotropically, has the
following Lagrangian:
\beq
{\cal L}=
{{\psi}^\dagger} \left( i\hbar\partial_t + \frac{\hbar^2}{2M}\nabla^2 \right) \psi
-\frac{C_0}{4} ({{\psi}^\dagger} \psi)^2
- \frac{C_2}{8} \nabla({{\psi}^\dagger } \psi)\nabla({{\psi}^\dagger } \psi) 
- \frac{D_0}{36} ({{\psi}^\dagger} \psi)^3\ +\ \ldots
\label{eq:1}
\eeq where the field operator $\psi$ destroys a boson.  The operators in
this Lagrangian are constrained by Galilean invariance, parity and
time-reversal invariance, and describe bosons which interact at
low-energies via an arbitrary potential of finite range.  The $C_{2n}$
are coefficients of two-body operators and $D_{0}$ is the coefficient
of a three-body operator.  This Lagrangian is valid in
any number of spacetime dimensions, $d$. The mass dimensions of the
boson field and of the operator coefficients depend on $d$ as follows:
$[\psi]=(d-1)/2$, $[C_{2n}]=2-d-2n$ and $[D_{0}]=3-2d$.  In this paper
bosons living in $d=4,3$ and $2$ spacetime
dimensions will be considered. While the coefficients of the operators are $d$-dependent
there is no need to label them as they are not observable quantities
and therefore no ambiguity will be encountered. By contrast, as will be seen, the
S-matrix takes a distinct form for each spacetime dimension.  Units
with $\hbar =1$ are used throughout and the boson mass, $M$, is kept
explicit.

\begin{figure}[!t]
  \centering
     \includegraphics[scale=0.5]{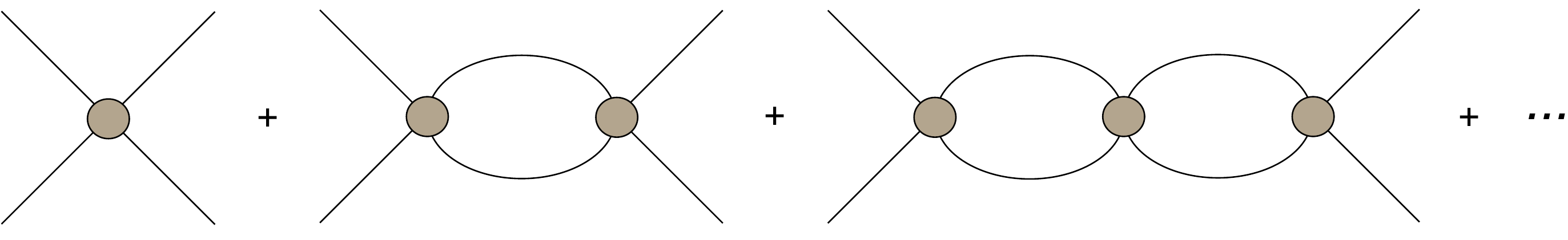}
     \caption{A diagrammatic representation of the scattering Amplitude ${\cal A}_2(p)$. The loop diagram with vertices removed is $I_0(p)$.}
  \label{fig:loops}
\end{figure}

Consider $2\rightarrow 2$ scattering, with incoming momenta
labeled ${\bf p}_1,{\bf p}_2$ and outgoing momenta labeled ${\bf
  p}'_1,{\bf p}'_2$. In the center-of-mass frame, ${\bf p}={\bf
  p}_1=-{\bf p}_2$ , and the sum of Feynman diagrams --illustrated in
Fig.~\ref{fig:loops}-- computed in the EFT gives the
two-body scattering amplitude
\begin{eqnarray}
{\cal A}_2(p) & = & -{ \sum C_{2n} \ p^{2n}  \over
1 - I_0(p) \sum C_{2n} \ p^{2n}} \ ,
\label{eq:2}
\end{eqnarray}
where
\begin{eqnarray}
I_0(p) \ = \ \frac{M}{2}\left({\mu\over 2}\right)^{\epsilon} \int {d^{D-1}{\bf q}\over
  (2\pi)^{D-1}}
{1\over p^2-{{\bf q}^2} + i \delta}
\ .
\label{eq:2b}
\end{eqnarray}
It is understood that the ultraviolet divergences in the EFT
are regulated using DR\footnote{In particular, the separation of the potential from the loop integral
in eq.(\ref{eq:2}) relies on special properties of DR, and would not, for instance, hold generally using cutoff regularization.}.
In eq.~(\ref{eq:2b}), $\mu$ and $D$ are the DR scale and dimensionality, respectively,
and $\epsilon\equiv d-D$. A useful integral is:
\begin{eqnarray}
\openup3\jot
I_n(p)&=& \frac{M}{2}\left({\mu\over 2}\right)^{\epsilon} \int {{{\rm d}}^{D-1}{\bf  q}\over (2\pi)^{D-1}}\, 
{\bf q}^{2n} \left({1\over p^2  -{\bf q}^2 + i\delta}\right) \ ;
\nonumber\\
&=& -\frac{M}{2} p^{2n} (-p^2-i\delta)^{(D-3)/2 } \Gamma\left({3-D\over 2}\right)
{(\mu/2)^{\epsilon}\over  (4\pi)^{(D-1)/2}}\ .
\label{eq:2a}
\end{eqnarray}
In this paper, the EFT coefficients will be defined in DR with $\overline{MS}$. This choice is 
convenient if the renormalized EFT coefficients are of natural size with respect to the 
distance scale $\ell$, which characterizes the range of the interaction. In systems with a
scattering length in three spatial dimensions which is large compared to $\ell$, it is convenient
to use DR with the $PDS$ scheme~\cite{Kaplan:1998tg}, which keeps the renormalized coefficients
of natural size in the presence of a large scattering length. However, it is important to emphasize
that there is no barrier to working in $\overline{MS}$ for a scattering length of any size as physics
is independent of the regularization and renormalization scheme.

The scattering amplitude can be parametrized via~\cite{Beane:2010ny}
\beq
{\cal A}_2(p) \ = \ \frac{-1}{{\rm Im}(I_0(p))\big\lbrack\cot\delta(p)-i\big\rbrack} \ ,
\label{eq:4}
\eeq
with
\beq
\cot\delta(p) \ = \ \frac{1}{{\rm Im}(I_0(p))}\Bigg\lbrack\frac{1}{\sum C_{2n} \ p^{2n}}\ -\  {\rm Re}(I_0(p)) \Bigg\rbrack \ .
\label{eq:5}
\eeq 
Bound states are present if there are poles of the scattering
amplitude on the positive, imaginary, momentum axis. That is, if
$\cot\delta(i\gamma)=i$ with binding momentum satisfying $\gamma >0$. 
Evaluating $I_0(p)$ in DR, it is convenient to consider even- and odd-spacetime dimensions separately.
For $d$ even the Gamma function has no poles and one finds the finite result
\begin{eqnarray}
I_0(p)&=& -\frac{M}{2(4\pi)^{(d-1)/2}}\frac{\pi i\; p^{d-3}}{\Gamma\left({d-1\over 2}\right)}\ .
\label{eq:Ievend}
\end{eqnarray}
Hence the $\overline{MS}$ EFT coefficients do not run with $\mu$ in
even spacetime dimensions and the bare parameters are renormalized parameters. For $d$ odd, one finds
\beq
I_0(p) =  \frac{M}{2(4\pi)^{(d-1)/2}}\frac{p^{d-3}}{\Gamma\left({d-1\over 2}\right)}
 \Bigg\lbrack \ln{\left(-\frac{p^2}{\mu^2}\right)}\ -\ \psi_0\left({d-1\over 2}\right)\ - \ln\pi\ -\ \frac{2}{\epsilon} \Bigg\rbrack \ ,
\label{eq:Iodd}
\eeq
where $\psi_0(n)$ is the digamma function. Here there is a single logarithmic divergence which is hidden in the $1/\epsilon$ pole.
Therefore in this scheme, at least one EFT coefficient will depend on the scale $\mu$. The general expression for the
isotropic phase shift in $d$ spacetime dimensions is:
\beq
p^{d-3}\cot\delta(p) \ = \  -\frac{(4\pi)^{(d-1)/2}}{\pi M}\Gamma\left({d-1\over 2}\right)\frac{2}{\sum C_{2n}\ p^{2n}}\ +\ (1-(-1)^d)\frac{p^{d-3}}{2\pi}\ln{\left(\frac{p^2}{\overline{\mu}^2}\right)}\ ,
\label{eq:gencotdelta}
\eeq
where $\overline{\mu}$ is defined by equating the logarithm in eq.~(\ref{eq:gencotdelta}) with the content of the square
brackets in eq.~(\ref{eq:Iodd}). This is, of course, an unrenormalized equation as the $C_{2n}$ coefficients are bare parameters
and there is a logarithmic divergence for odd spacetime dimensions. 

\section{Isotropic scattering in the continuum}
\label{sec:IsoSCatt}

\subsection{$d=2$: one spatial dimension}

\noindent In one spatial dimension, eq.~(\ref{eq:gencotdelta}) gives
\begin{eqnarray}
p^{-1}\cot\delta(p) \ = \ - a_{1}\ +\ \tau_1\, p^2 \ + \ \sum_{n=2}^\infty u_{(n)}p^{2n}
\label{eq:7}
\end{eqnarray}
with scattering length and volume, respectively,
\begin{eqnarray}
a_{1}\ =\ \frac{4}{M C_0}\ ; \qquad  \tau_1\ =\ \frac{4C_2}{M C_0^2}\ .
\label{eq:7b}
\end{eqnarray}
The $u_{(n)}$ are shape parameters which are easily matched to the $C_{2n}$ coefficients. Neglecting the scattering
volume, for $a_{1}<0$ there is a bound state with binding momentum $\gamma_1=-1/a_{1}$.

\subsection{$d=4$: three spatial dimensions}

\noindent For three spatial dimensions, eq.~(\ref{eq:gencotdelta}), yields the familiar effective range expansion,
\begin{eqnarray}
p\cot\delta(p) \ = \ -\frac{1}{a_{3}}\ +\ r_{3}\, p^2 \ + \ \sum_{n=2}^\infty v_{(n)}p^{2n}
\label{eq:7c}
\end{eqnarray}
with scattering length\footnote{Note that in the $PDS$ scheme~\cite{Kaplan:1998tg}, the relationship between the scattering length and the
renormalized coefficient is modified to 
\begin{eqnarray}
\frac{8\pi}{M C_0(\mu)}=\frac{1}{a}-\mu
\label{eq:7cmod}
\end{eqnarray}
where $\mu$ is the $PDS$ renormalization scale. The $\overline{MS}$ scheme is recovered as $\mu\rightarrow 0$.
} 
and effective range, respectively,
\begin{eqnarray}
a_{3}\ =\ \frac{M C_0}{8\pi}\ ; \qquad  r_{3}\ =\ \frac{16\pi C_2}{M C_0^2}\ .
\label{eq:7d}
\end{eqnarray}
The $v_{(n)}$ are shape parameters.
Neglecting the effective range, for $a_{3}>0$ there is a bound state with binding momentum $\gamma_3=1/a_{3}$.

\subsection{$d=3$: two spatial dimensions}

\noindent In this section, the case $d=3$ will be considered in some detail. 
From the general formula, eq.~(\ref{eq:gencotdelta}), one finds
\begin{eqnarray}
\cot\delta(p) \ = \ \frac{1}{\pi} \ln{\left(\frac{p^2}{\mu^2}\right)} \ -\ \frac{1}{\alpha_{2}(\mu)}\ +\ \sigma_2\, p^2 \ + \ \sum_{n=2}^\infty w_{(n)}p^{2n}
\label{eq:9}
\end{eqnarray}
with coupling constant and effective area, respectively,
\begin{eqnarray}
\alpha_{2}(\mu )\ =\ \frac{M C_0(\mu )}{8}\ ; \qquad  \sigma_2\ =\ \frac{8C_2(\mu)}{M C_0^2(\mu )} \ .
\label{eq:9b}
\end{eqnarray}
Note that $\sqrt{|\sigma_2|}$ is the effective range.  The $w_{(n)}$
are shape parameters.  Neglecting all higher-order range corrections,
for $\alpha_{2}(\mu )$ of either sign, there is a bound state with
binding momentum $\gamma_2=\mu \exp(\pi/2\alpha_{2}(\mu ))$. This
occurs because quantum mechanical effects generate an attractive
logarithmic contribution which always dominates at long
distances. However, in the repulsive case this bound state is not
physical.

The scale dependence of the leading EFT coefficient is determined by the
condition that the scattering amplitude be independent of the scale $\mu$:
\begin{eqnarray}
\mu\frac{d}{d\mu}C_0(\mu )\ = \ \frac{M}{4\pi}{C_0^2}(\mu )\ .
\label{eq:10a}
\end{eqnarray}
This equation is readily integrated to give the renormalization group evolution equation
\begin{eqnarray}
\alpha_2(\mu )\ = \ \frac{\alpha_2(\nu)}{1- \frac{2}{\pi}\alpha_2(\nu )\ln\left(\frac{\mu}{\nu}\right)} \ .
\label{eq:rg1}
\end{eqnarray}
It is clear from eq.~(\ref{eq:rg1}) that the attractive case,
$\alpha_2(\mu )=-|\alpha_2(\mu )|$, corresponds to an asymptotically
free coupling, while the repulsive case, $\alpha_2(\mu
)=+|\alpha_2(\mu )|$, has a Landau pole and the coupling grows weaker
in the infrared. The position of the bound state in the repulsive case
is the position of the Landau pole, which sets the cutoff scale of the
EFT; this is the scale at which new ultraviolet physics should make
its appearance and effectively remove the singularity. Therefore, the 
bound state in the repulsive case is unphysical.

A more common parametrization of the phase shift is given by
\begin{eqnarray}
\cot\delta(p) \ = \ \frac{1}{\pi} \ln{\left({p^2}{a_2^2}\right)} \ +\ \sigma_2\, p^2 \ + \ \sum_{n=2}^\infty w_{(n)}p^{2n} \ ,
\label{eq:9mod}
\end{eqnarray}
where $a_2$ is the scattering length in two spatial dimensions. By matching with eq.~(\ref{eq:9}), one
finds $a_2^{-1}=\mu \exp(\pi/2\alpha_{2}(\mu ))$, which in the repulsive case is the position of the
Landau pole. Hence, in the repulsive case, $a_2^{-1}$ is the momentum cutoff scale. 
This suggests that $a_2$ is not an optimal parameter for describing low-energy
physics since it is unnatural with respect to the characteristic interaction length scale, $\ell$.
By contrast, one expects that the dimensionless parameter $\alpha_2(\mu )$ will take a natural
size when $\mu\sim \ell^{-1}$.

\section{Compactification}
\label{sec:compact}

\begin{figure}[!t]
  \centering
     \includegraphics[scale=0.55]{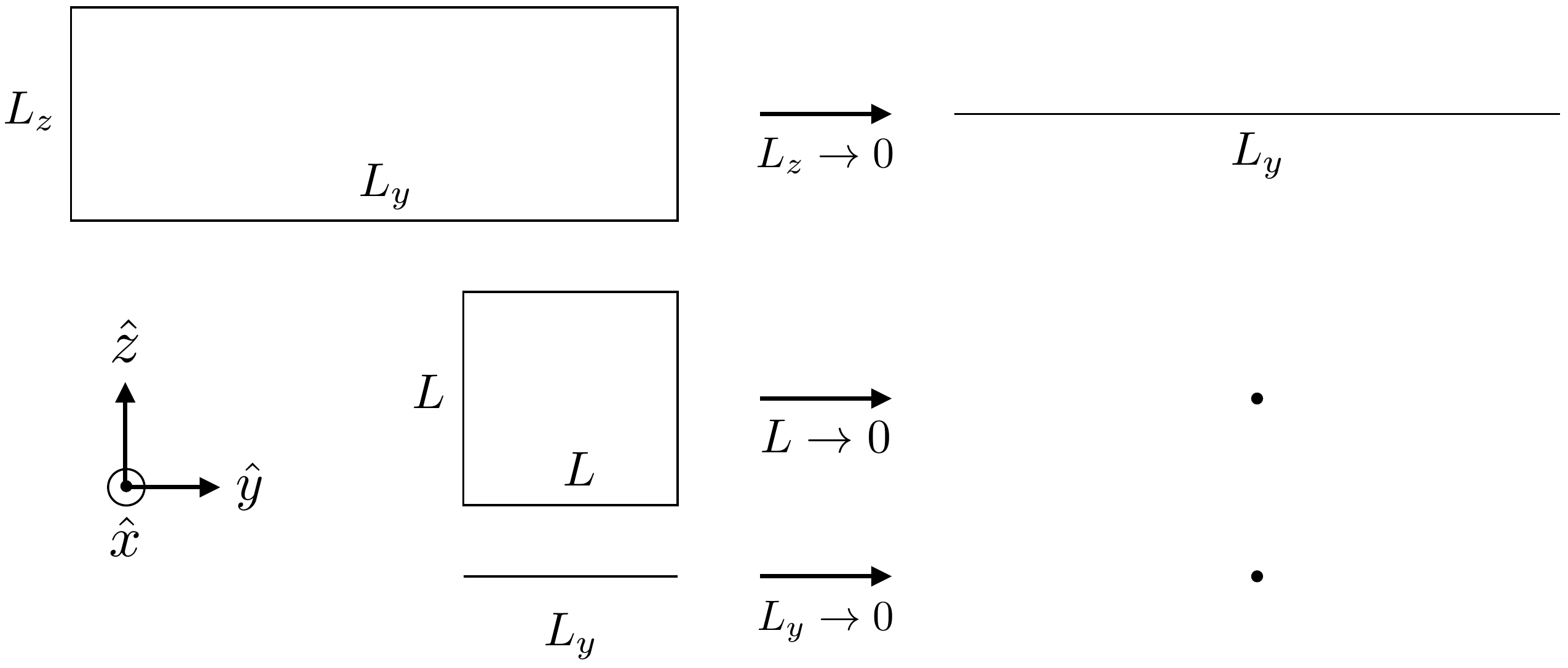}
     \caption{Patterns of compactification considered in the text. The ${\hat x}$-direction, which points out of the page, is always taken to be infinite.}
  \label{fig:compact}
\end{figure}

\noindent In this section, toroidal compactifications of space, which
interpolate between the various cases at fixed spatial dimension
outlined above, will be considered. The procedure is simple and
intuitive.  In the two-body scattering problem, all the effects that
arise from placing an infrared boundary on a dimension, appear through
the loop integral, $I_0(p)$.  If the characteristic range of the
interaction is taken to be $\ell$, then placing the two-body system in
a box of sides $\{L_x,L_y,L_z\}$ with periodic boundary conditions
quantizes the momenta that can run around the loop such that
\begin{eqnarray}
q_i&=&  \frac{2\pi n_i}{L_i}
\label{eq:qunmom}
\end{eqnarray}
where ${\bf n}\in\mathbb{Z}^{d-1}=(n_x,n_y,\ldots,n_{d-1})$.
In Fig.~\ref{fig:compact} the various patterns
of compactification considered here are illustrated\footnote{An alternative means
of moving continuously among various dimensions is to constrain two-particle scattering
to a cylinder of radius $R$ and to consider the limit of a plane ($R\rightarrow\infty$)
and of a wire ($R\rightarrow 0$) as is relevant in the case of a carbon nanotube~\cite{Delfino:2011a}.}
in order of consideration. The 
distinct regimes depend on the ratio of the size of the
``compactified'' dimension and the physical scales of the problem. 
If $L_x,L_y,L_z\gg \ell$ then the continuum results are recovered.

\subsection{$d=4$ to $d=3$}
\label{sec:compact43}

\noindent Consider the case of three spatial dimensions with one
dimension compactified on a circle. One has a box with sides
$\{L_x,L_y,L_z\}$ and $L_x,L_y\gg \ell$ where again $\ell$ represents the
characteristic range of the interaction.  Near the continuum limit in
the $x$- and $y$-directions, and assuming that $L_z$ is finite, the
topology of space is $\mathbb{R}^2\times \mathbb{S}^1$.  Hence, one can
write
\begin{eqnarray}
I^{L_z}_0(p) \ =\ \frac{M}{2}\frac{1}{L_z}\sum_{q_z} \int \frac{d^2q}{(2\pi)^2}{1\over p^2-q_z^2-q^2+i\delta} \ ,
\label{eq:compact1}
\end{eqnarray}
where here $q^2\equiv q_x^2+q_y^2$. Noting that putting the infrared boundary in the ${\hat z}$-direction has not altered the ultraviolet
behavior, evaluating this expression in $\overline{MS}$ yields
\begin{eqnarray}
I^{L_z}_0(p) \ =\ \frac{M}{4\pi L_z}\, \ln\left( 2 e^{-i\frac{\pi}{2}} \sin\left(\frac{L_z}{2}\sqrt{p^2 + i\delta}\right)\right) \ .
\label{eq:compact2}
\end{eqnarray}
In the continuum limit, $L_z\rightarrow\infty$, and tracking the parameter $\delta$, one finds
\begin{eqnarray}
I^{\infty}_0(p) \ =\ -\frac{i M p}{8\pi}
\label{eq:compact3}
\end{eqnarray}
as expected from eq.~(\ref{eq:Ievend}). When $L_z p\ll 1$ , separating off the non-analytic piece yields
\begin{eqnarray}
I^{L_z}_0(p) \ =\ \frac{M}{4\pi L_z} \left(\ -\frac{i\pi}{2}\ +\ \ln\left(L_z p\right)\ -\ \sum_{n=1}^\infty \frac{1}{n}\left( \frac{L_z p}{2\pi} \right)^{2n}{\bf \zeta}(2n) \right) \ ,
\label{eq:compact4}
\end{eqnarray}
where ${\bf \zeta}(s)$ is the Riemann zeta function. 
In this limit of scattering in three spatial dimensions eq.~(\ref{eq:5}) gives
\begin{eqnarray}
\cot\delta(p) \ = \ -\frac{8 L_z}{M}\Bigg\lbrack \frac{1}{C_{0}}\ - \ \frac{C_2}{C_0^2}p^2   \ -\  \frac{M}{4\pi L_z} \left(\ \ln\left(L_z p\right)\ -\ \frac{L_z^2 p^2}{24} \right) \ +\ {\mathcal O}(p^4)\  \Bigg\rbrack \ ,
\label{eq:compact5}
\end{eqnarray}
where the $C_{2n}$ are bare $d=4$ coefficients.
Using eq.~(\ref{eq:7d}) to express the EFT coefficients in terms of the effective range parameters in three spatial dimensions
and matching to the general form for the phase shift in two spatial dimensions, eq.~(\ref{eq:9mod}), yields\footnote{Note that in the $PDS$ scheme,
eq.(\ref{eq:compact2}) has an additional piece, $-M\mu/8\pi$, which then gives the correct relationship between the scattering length and the
renormalized coefficient in eq.(\ref{eq:compact5}) so that all relations involving the effective range parameters are independent of the renormalization scheme.}
\begin{eqnarray}
a_2 \ =\ L_z \exp\left({-\frac{L_z}{2a_3}}\right) \ ; \qquad \sigma_2 \ = \ \frac{L_z}{2\pi}\left(r_3\ -\ \frac{L_z}{6} \right) \ .
\label{eq:compact6}
\end{eqnarray}
The expression matching the scattering lengths is in agreement with previous results found in Refs.~\cite{PhysRevA.93.063631,2018arXiv180601784I}.
The matching of higher-order effective range parameters gives (for $n\geq 2$)
\begin{eqnarray}
w_{(n)} &=& \frac{L_z}{\pi} v_{(n)} \;-\; \frac{2}{\pi n}\left(\frac{L_z}{2\pi} \right)^{2n} {\bf \zeta}(2n) \ .
\label{eq:compact6ap}
\end{eqnarray}

The condition for a bound state in the presence of the boundary is
\begin{eqnarray}
1 \ =\ 2\sinh\left(\frac{L_z \gamma_2}{2}\right) \exp\left({-\frac{L_z}{2}}\left(\frac{1}{a_3}+...  \ldots\right)\right) 
\label{eq:compactBSgen}
\end{eqnarray}
where $\gamma_2$ is the binding momentum and the dots signify higher order terms in the effective range expansion.
Hence, in the limit of two spatial dimensions where $L_z\rightarrow 0$ the binding momentum
tends to the universal value
\begin{eqnarray}
\gamma_2 &=& \ln \left(\oneht\left(3+\sqrt{5}  \right)  \right) L_z^{-1} \; =\; \left(0.962423650\ldots \right) L_z^{-1} \ .
\label{eq:compact7be}
\end{eqnarray}
This value is independent of details of the finite-range potential in
three spatial dimensions\footnote{This point has been made previously in Ref.~\cite{0953-4075-48-2-025302}.}. For instance, even if the two-body
attraction in the three-dimensional theory is not enough for binding,
the presence of the boundary and the compactification to two
dimensions binds the system. This realizes in practice for this
particular geometry the observation made in Sec.~\ref{sec:IsoSCatt}
that there is always a bound state in two dimensions due to strong
infrared effects.

If there is a bound state in the three-dimensional theory with binding
momentum $\gamma_3$, then in the presence of the boundary the binding
momentum is given by the approximate formula
\begin{eqnarray}
\gamma_2 \ =\ \frac{2}{L_z}\ln\left( \oneht \exp\left(\frac{L_z\gamma_3}{2}\right)+\sqrt{1+\onefourth\exp\left({L_z\gamma_3}\right)}   \right) \ ,
\label{eq:compactLogform}
\end{eqnarray}
which smoothly interpolates between the universal value of
eq.~(\ref{eq:compact7be}) in the limit of two spatial dimensions and
the binding momentum in the box as the boundary is removed. If there
is no bound state in the three-dimensional theory, then the appearance
of the boundary signals the presence of a bound state at threshold
which again tends toward the universal value of
eq.~(\ref{eq:compact7be}) in the limit of two spatial dimensions.

\begin{figure}[!t]
  \centering
     \includegraphics[scale=0.38]{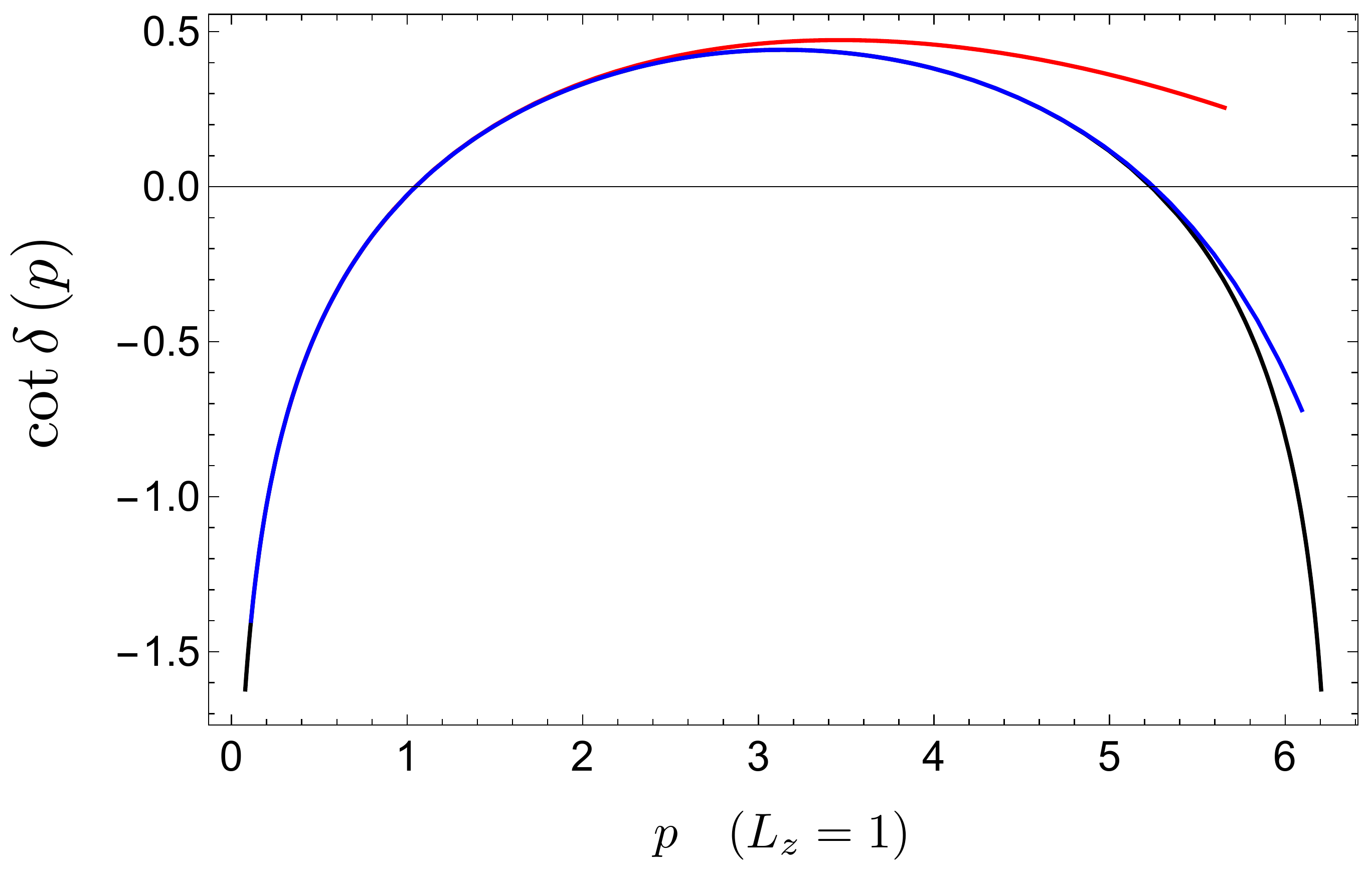}
     \caption{Exact S-matrix in the presence of a boundary when two-body system is at unitarity in $d=4$. Units are chosen with $L_z=1$. The black curve is the exact phase shift, the red
curve is the low-momentum expansion up to effective area corrections, and the blue curve includes eight orders in the effective range expansion.}
  \label{fig:unitarity43}
\end{figure}

Now consider the special case where the original theory in three
spatial dimensions is at or near unitarity; this corresponds to
$a_3\rightarrow\infty$, while $r_3,v_{(n)}\rightarrow 0$\footnote{See, for
  instance, Ref.\cite{Kolck:2017zzf}.}.  In this case, the original
(two-body) theory is at a non-trivial fixed point of the
renormalization group and therefore has a non-relativistic conformal
invariance, i.e. Schr\"odinger symmetry. One consequence of this
symmetry is that the bound state in the system has zero energy (as there
is no scale). The exact phase shift in the presence of the boundary is
then given by~\cite{0953-4075-48-2-025302}
\begin{eqnarray}
\cot\delta(p) \ = \ \frac{2}{\pi}\, \ln\left( 2 \sin\left(\frac{L_z p}{2}\right)\right) \ ,
\label{eq:compact7}
\end{eqnarray}
for $L_z p\sim 1$, with the restriction $0< {L_z p}/{2}<\pi$. This phase shift is plotted in
Fig.~\ref{fig:unitarity43} and compared with effective range theory.

As the boundary is brought in from infinity where there is a bound
state at threshold, one expects that the binding momentum should scale
as $1/L_z$, and indeed one finds from eq.~(\ref{eq:compact7}) (and
eq.~(\ref{eq:compactBSgen})) that the binding momentum is given by
eq.~(\ref{eq:compact7be}).  In the limit of two spatial dimensions
where $L_z\rightarrow 0$, all effective range parameters, and
therefore the EFT, is entirely fixed by the one parameter, $L_z$,
which breaks the scale invariance.  Evidently this two-dimensional
theory is repulsive and indeed in this limit the binding energy rises
to its largest possible value which is of course the position of the
Landau pole.  Hence this EFT in two spatial dimensions with repulsive
interactions has as its ultraviolet completion (for $p\gg L_z^{-1}$) a conformal field
theory in three spatial dimensions.

\subsection{$d=4$ to $d=2$}
\label{sec:compact42}

\noindent Now consider the case of three spatial dimensions
with two dimensions compactified on a sphere. Take a box with
sides $\{L_x,L_y,L_z\}$ and choose $L_x\gg \ell$. 
Near the continuum limit in the $x$-direction, and assuming that $L_z=L_y\equiv L$ is finite, 
the topology of space is $\mathbb{R}^1\times \mathbb{S}^2$.
Hence, the loop integral becomes
\begin{eqnarray}
I^{L^2}_0(p) \ =\ \frac{M}{2}\frac{1}{L^2}\sum_{q_y,q_z} \int_{-\infty}^{\infty} \frac{dq_x}{(2\pi)}{1\over p^2-q_x^2-q_y^2-q_z^2+i\delta} \ .
\label{eq:compact421}
\end{eqnarray}
Evaluating the integral in $\overline{MS}$ yields
\begin{eqnarray}
I^{L^2}_0(p) \ =\ -\frac{M}{4L^2}\Bigg\lbrack \frac{i}{p}\;+\;\frac{L}{2\pi} \left( \sum^{\Lambda_n}_{n_y,n_z\neq 0} \frac{1}{\sqrt{n_y^2+n_z^2-{\tilde p}^2}}-2\pi\Lambda_n\right) \Bigg\rbrack\ ,
\label{eq:compact422c}
\end{eqnarray}
where ${\tilde p}\equiv  p L/2\pi$, and $\Lambda_n\rightarrow\infty$ in an integer cutoff. This two-dimensional sum is tractable in the sense that it can be expressed as a one-dimensional
sum over special functions. It is straightforward to find~\cite{Glasser1,Beane:2010ny}
\begin{eqnarray}
\sum^{\Lambda_n}_{n_y,n_z\neq 0} \frac{1}{\sqrt{n_y^2+n_z^2-{\tilde p}^2}}-2\pi\Lambda_n &=& \sum^{\infty}_{k= 0}(-1)^k\Choose{-\oneht}{k}{\tilde p}^{2k}\; 4{\bf \zeta}(\oneht+k){\bf \beta}(\oneht+k) \ ,
\label{eq:compact423}
\end{eqnarray}
where ${\bf \beta}(s)$ is the Dirichlet beta function\footnote{Equivalently, one can express the two-dimensional sum as a one-dimensional sum over a single special function
\begin{eqnarray}
\sum^{\Lambda_n}_{n_y,n_z\neq 0} \frac{1}{\sqrt{n_y^2+n_z^2-{\tilde p}^2}}-2\pi\Lambda_n &=& 4\sum^{\infty}_{k= 0}\frac{(-1)^k}{\sqrt{(2k+1)}} {\bf \zeta}\left(\oneht,1-\frac{{\tilde p}^2}{(2k+1)}\right)
\label{eq:compact424}
\end{eqnarray}
where ${\bf \zeta}(s,a)$ is the Hurwitz zeta function.}. The phase shift is now given by
\begin{eqnarray}
\!\!\!\!\!\!p^{-1}\cot\delta(p) = -\frac{4L^2}{M}\Bigg\lbrack \frac{1}{C_{0}}\; - \; \frac{C_2}{C_0^2}p^2   \; +\;  \frac{M}{8\pi L} \left(  4{\bf \zeta}(\oneht){\bf \beta}(\oneht)\;+\; \frac{L^2}{2\pi^2}{\bf \zeta}(\threeht){\bf \beta}(\threeht)p^2  \right)  + {\mathcal O}(p^4) \Bigg\rbrack 
\label{eq:compact42phase}
\end{eqnarray}
where again the $C_{2n}$ are bare $d=4$ coefficients. Proceeding as before one finds
\begin{eqnarray}
a_1 \ =\ \frac{L^2}{2\pi} \left( \frac{1}{a_3}\;+\; \frac{1}{L}4{\bf \zeta}(\oneht){\bf \beta}(\oneht) \right) \ \ \ ,\ \ \ \tau_1 \ =\  \frac{L^2}{4\pi} \left( {r_3}\;-\; \frac{L}{\pi^2}{\bf \zeta}(\threeht){\bf \beta}(\threeht) \right) \ ,
\label{eq:compact6aq}
\end{eqnarray}
where
\begin{eqnarray}
4{\bf \zeta}(\oneht){\bf \beta}(\oneht) &=&  -3.9002649200019558828454753366049732192090478564775  \ ;   \nn \\
{\bf \zeta}(\threeht){\bf \beta}(\threeht)  &=& 2.2584054207752375764326288198292646348757913461436    \ .
\label{eq:compact6ar}
\end{eqnarray}
The reason for quoting so many digits will be made clear below. 
The expression matching the scattering lengths is in agreement with the result found in Ref.~\cite{2018arXiv180601784I}\footnote{Note that Ref.~\cite{2018arXiv180601784I} uses the opposite
sign convention for the one-dimensional scattering length.}.
It is straightforward to match higher-order effective range parameters giving (for $n\geq 2$)
\begin{eqnarray}
u_{(n)} &=& \frac{L^2}{2\pi} v_{(n)} \;-\; (-1)^n\Choose{-\oneht}{n}\left( \frac{L}{2\pi}\right)^{2n+1}\; 4{\bf \zeta}(\oneht+n){\bf \beta}(\oneht+n)  \ .
\label{eq:compact6b}
\end{eqnarray}
The one-dimensional limit, $L\rightarrow 0$, supports a bound state with binding momentum determined by the roots of
\begin{eqnarray}
-1 \ = \ \sum^{\infty}_{k= 0}\Choose{-\oneht}{k} \left( \frac{L \gamma_1}{2\pi}\right)^{2k+1}\; 4{\bf \zeta}(\oneht+k){\bf \beta}(\oneht+k) \ .
\label{eq:compact7az}
\end{eqnarray}
One finds a bound state with binding momentum
\begin{eqnarray}
\gamma_1 = \left(1.511955584\ldots \right) L^{-1} \ .
\label{eq:compact7azba}
\end{eqnarray}
As in the previous case, this result is universal in the sense that it is independent of whether the initial system is bound.

\begin{figure}[!t]
  \centering
     \includegraphics[scale=0.38]{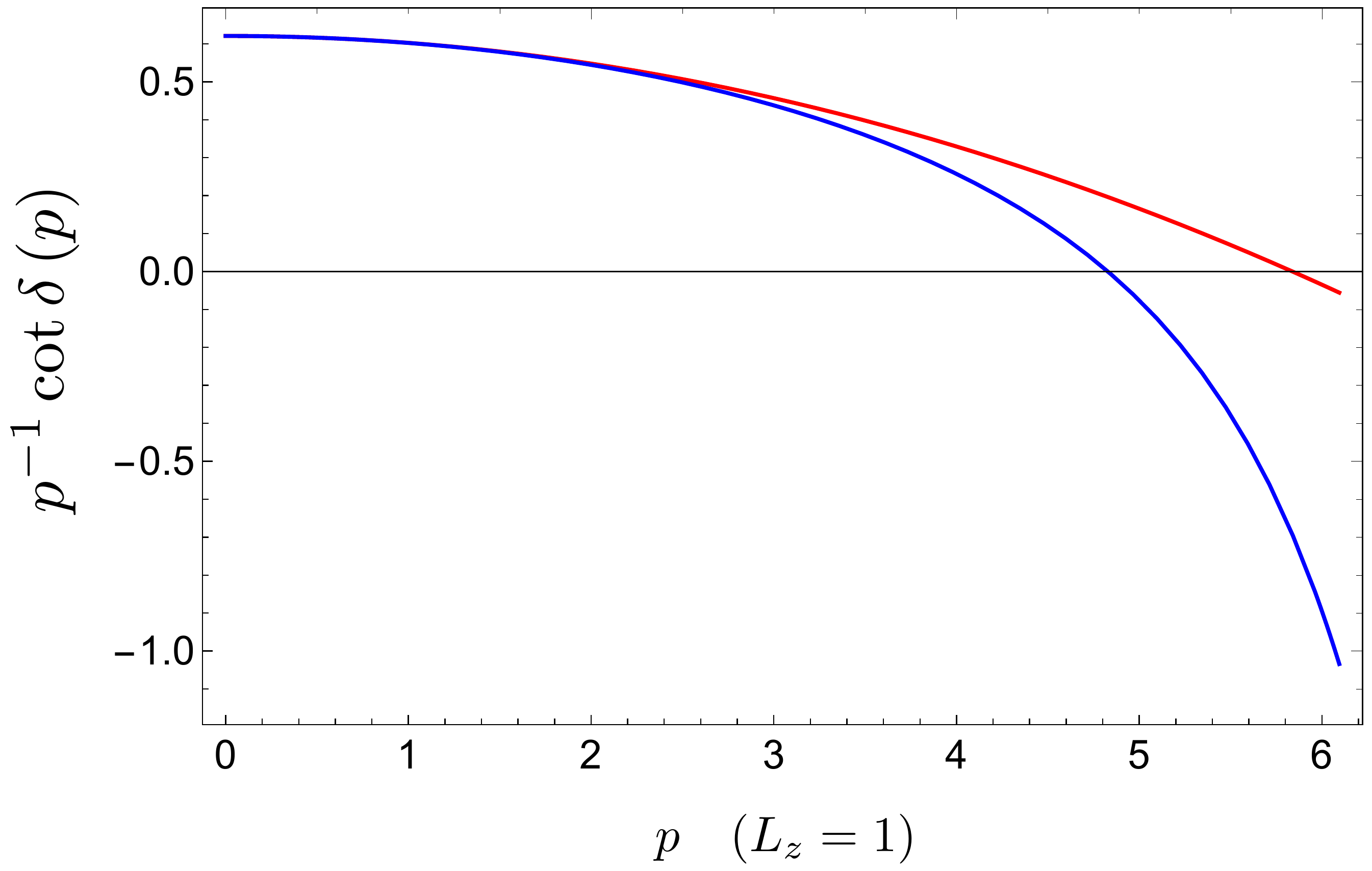}
     \caption{Exact S-matrix in the presence of a boundary when two-body system is at unitarity in $d=4$. Units are chosen with $L=1$. The red
curve is the low-momentum expansion up to effective area corrections, and the blue curve includes eight orders in the effective range expansion.}
  \label{fig:unitarity42}
\end{figure}

One can again consider the special case where the original theory in three spatial dimensions is at
unitarity. Here the exact phase shift in the one-dimensional theory with compactified dimensions is
\begin{eqnarray}
p^{-1}\cot\delta(p) \ = \ -\left( \frac{L}{2\pi}\right)\sum^{\infty}_{k= 0}(-1)^k\Choose{-\oneht}{k} \left( \frac{L p}{2\pi}\right)^{2k}\; 4{\bf \zeta}(\oneht+k){\bf \beta}(\oneht+k) \ ,
\label{eq:compact7a}
\end{eqnarray}
which supports a bound state with binding momentum given by eq.~(\ref{eq:compact7azba}). This phase shift is plotted in Fig.~\ref{fig:unitarity42} and compared with effective range theory.

\subsection{$d=3$ to $d=2$}
\label{sec:compact32}

\noindent Lastly consider the case of two spatial dimensions
with one dimension compactified on a circle. Take a square with
sides $\{L_x,L_y\}$ and choose $L_x\gg \ell$. 
Near the continuum limit in the $x$-direction, and assuming that $L_y$ is finite, 
the topology of space is $\mathbb{R}^1\times \mathbb{S}^1$.
Hence, the loop integral becomes
\begin{eqnarray}
I^{L_y}_0(p) \ =\ \frac{M}{2}\frac{1}{L_y}\sum_{q_y} \int_{-\infty}^{\infty} \frac{dq_x}{(2\pi)}{1\over p^2-q_x^2-q_y^2+i\delta} \ .
\label{eq:compact421a}
\end{eqnarray}
This case is somewhat more involved than the other cases as there is a logarithmic divergence which is not affected
by the infrared boundary. As the integral is linearly convergent, this may be evaluated first, giving
\begin{eqnarray}
I^{L_y}_0(p) \ =\ -\frac{M}{4 L_y}\left( \frac{i}{p}\;+\; \sum_{q_y\neq 0}^{\overline{MS}}  \frac{1}{\sqrt{q_y^2-p^2}}\right) \ ,
\label{eq:compact422}
\end{eqnarray}
where the logarithmically-divergent sum must be evaluated in $\overline{MS}$. One finds
\begin{eqnarray}
\!\!\!\!\!\sum_{q_y\neq 0}^{\overline{MS}}  \frac{1}{\sqrt{q_y^2-p^2}} = 
\frac{L_y}{2\pi}\Bigg\lbrack 2\left(  \sum_{n_y= 1}^{\Lambda_n} \frac{1}{\sqrt{n_y^2-{\tilde p}^2}}-\ln\Lambda_n \right)+ 2\ln\left(\frac{\mu L_y}{4\pi} \right)
-\gamma_E+\ln\pi+\frac{2}{\epsilon} \Bigg\rbrack ,
\label{eq:compact422a}
\end{eqnarray}
where here ${\tilde p}\equiv  p L_y/2\pi$ and $\gamma_E=\psi_0(1)$ is the Euler-Mascheroni constant. Renormalizing the bare two-dimensional coefficients, as outlined in Sec.~\ref{sec:IsoSCatt}, it follows that 
\begin{eqnarray}
\cot\delta(p) & = & -\frac{L_y p}{2}\Bigg\lbrack \frac{1}{\alpha_{2}(\mu)}\; +\; \frac{2}{\pi}\ln\left(\frac{\mu L_y}{4\pi} \right) \; -\; \sigma_2\, p^2 \; - \; \sum_{n=2}^\infty w_{n}p^{2n} \nn \\
&&\qquad\qquad\qquad\;+\; \frac{2}{\pi}\left(  \sum_{n_y= 1}^{\Lambda_n} \frac{1}{\sqrt{n_y^2-{\tilde p}^2}}-\ln\Lambda_n \right) \Bigg\rbrack \ .
\label{eq:999}
\end{eqnarray}
Using the renormalization group equation, eq.~(\ref{eq:rg1}), then gives
\begin{eqnarray}
p^{-1}\cot\delta(p) = \frac{L_y}{2}\Bigg\lbrack -\frac{1}{\alpha_{2}\left(\frac{4\pi}{L_y}\right)}+ \sigma_2\, p^2 
- \frac{2}{\pi}\left( \gamma_E+\frac{L_y^2}{8\pi^2}{\bf \zeta}(3)p^2 \right)  + {\mathcal O}(p^4) \Bigg\rbrack.
\label{eq:9999}
\end{eqnarray}
In terms of the two-dimensional scattering length, $a_2$, one then finds
\begin{eqnarray}
a_1 &=& \frac{L_y}{\pi} \ln\left(\frac{L_y e^{\gamma_E}}{4\pi a_2} \right) \ \ \ , \ \ \  
\tau_1 \ =\  \frac{L_y}{2}\sigma_2\;-\; \left(\frac{L_y}{2\pi}\right)^3 {\bf \zeta}(3) \ ,
\label{eq:99999}
\end{eqnarray}
and
\begin{eqnarray}
u_{(n)} &=& \frac{L_y}{2} w_{(n)} \;-\; \frac{L_y}{\pi}(-1)^n\Choose{-\oneht}{n}\left( \frac{L_y}{2\pi}\right)^{2n}\;{\bf \zeta}(2n+1) \ .
\label{eq:compact6bb}
\end{eqnarray}
Note that in the one-dimensional limit, there is again always a bound state approaching threshold. However, the bound state that is always present in the initial two-dimensional theory
can be lost in the presence of the boundary only to reappear again when a bound state appears at threshold in the approach to the one-dimensional limit when $L_y\sim 4\pi a_2 e^{-\gamma_E}$.

\subsection{$d=4$ to $d=2$ in two steps}

\noindent Combining the results of Sec.~\ref{sec:compact43} and Sec.~\ref{sec:compact32}, the compactification from $d=4$ to $d=2$ can be achieved in two steps.
Setting $L_z=L_y=L$, one obtains
\begin{eqnarray}
\!\!\!a_1 \ =\ \frac{L^2}{2\pi} \left( \frac{1}{a_3}\;+\; \frac{1}{L}2\left(\gamma_E -\ln 4\pi  \right) \right) \ \ \ ,\ \ \ \tau_1 \ =\  \frac{L^2}{4\pi} \left( {r_3}\;-\; \frac{L}{\pi^2}\frac{1}{2}\left(
{\bf \zeta}(3)+\frac{\pi^2}{3}  \right) \right)  ,
\label{eq:compact6ab}
\end{eqnarray}
where
\begin{eqnarray}
2\left(\gamma_E -\ln 4\pi  \right) &=& -3.9076171641355158647427590083740188335122714913918  \ ;  \nn \\
\frac{1}{2}\left({\bf \zeta}(3)+\frac{\pi^2}{3}  \right) &=& 2.2459625184280235791722842474017501846014430473770    \ .
\label{eq:compact6abc}
\end{eqnarray}
Note that these expressions differ at the one-part-per-mil level from the expressions obtained in the one-step compactification, eq.~(\ref{eq:compact6ar}).
It is straightforward to match higher-order effective range parameters giving (for $n\geq 2$)
\begin{eqnarray}
u_{(n)} &=& \frac{L^2}{2\pi} v_{(n)} \;-\;     2\left(\frac{L}{2\pi}\right)^{2n+1}\left( \frac{1}{n}{\bf \zeta}(2n)\;+\; (-1)^n \Choose{-\oneht}{n}{\bf \zeta}(2n+1) \right)\ .
\label{eq:compact6bba}
\end{eqnarray}
This difference with the one-step compactification of
eq.~(\ref{eq:compact6b}) increases with $n$.  The discrepancy between
the one- and two-step compactifications to the wire is not surprising
as the initial geometries differ; in the two-step case, the initial
compactification to the plane assumed that the ${\hat y}$-direction
was infinite, whereas in the one-step case both the ${\hat y}$- and
${\hat z}$-directions were finite in extent and equal. The near
equality, particularly between the expressions
eq.~(\ref{eq:compact6ar}) and eq.~(\ref{eq:compact6abc}), is
intriguing given that there are no known expressions of the Riemann and
Dirichlet beta functions of half integer argument in terms of
fundamental constants.

\section{Conclusion}
\label{sec:Conc}

\noindent Interesting quantum mechanical phenomena, like resonance
effects that occur in few-body systems and phase transitions which
occur in many-body systems, depend critically on the strength and form
of the quantum mechanical potential and on environmental constraints
like temperature and spatial dimensionality. Given recent experimental
progress in controlling spatial dimensionality, it is of interest to
consider properties of general quantum mechanical systems as the spatial
dimensionality is varying. This paper has considered a very general
type of non-relativistic quantum mechanical system of bosons that
interact entirely via finite-range interactions. Starting from a world
with three spatial dimensions, it is straightforward to perform
toroidal compactifications to worlds with two and one spatial
dimensions. It is somewhat counter intuitive that in some sense the
most difficult aspect of this general problem is properly accounting
for the renormalization of ultraviolet divergences which are 
unaffected by the infrared boundaries that enter through the
compactification procedure. The use of dimensional regularization with
minimal subtraction greatly simplifies the computations, primarily
because power-law divergences do not appear in this scheme. General
relations among effective range parameters were obtained between
various dimensions. These relations may be useful in computing
non-universal corrections to Bose gas thermodynamic variables 
in various dimensional crossover schemes. For instance, the
effective range (area) corrections to the weakly interacting Bose gas
in two spatial dimension were recently computed in Ref.~\cite{Beane:2018jyq}.
That result, together with the expression for the effective range
given in eq.~(\ref{eq:compact6}), yields the leading non-universal
correction due to the effective range to the energy of the quasi-two
dimensional Bose gas given in Ref.~\cite{2018arXiv180601784I}.

An interesting consequence of compactification found in this paper 
is that even if the initial two-body system is not bound in three dimensions,
as the boundary is removed and the system is compactified to a plane
or a wire, the resulting two-body system always ends up bound.
This paper also considered the theoretical scenario 
where scattering in three spatial dimensions is at unitarity, and this
conformal system is compactified to a plane or to a wire. The resulting
S-matrices in the reduced dimensionality are then known exactly, and
are, of course, universal functions of the compactified length scale,
since the underlying theory has no scale. This provides an interesting
example of an EFT in two-spatial dimensions with repulsive interactions
where the presence of the Landau pole is traced to the underlying theory
which is given by a three dimensional system at unitarity.

Finally, it should be mentioned that the EFT methods used here to
obtain the relations among all effective range parameters for the case
of toroidal compactification can also be fruitfully applied to the
case of compactification achieved via the presence of atomic traps,
which effectively confine the particles using harmonic
potentials~\cite{0953-4075-51-6-065004}. In addition, the
consideration of three- and four-body systems as dimensionality is
altered~\cite{0953-4075-48-2-025302,0953-4075-51-6-065004} is also a
straightforward extention of EFT methods, although the analysis is
significantly more involved than in the two-body sector.

\section*{Acknowledgments}

\noindent We would like to thank Dmitry Petrov, Varese Tim\'oteo and Nikolaj Zinner for valuable comments on the manuscript.
This work was supported in part by the U.~S.~Department of Energy grant DE-SC001347.

\bibliographystyle{JHEP}
\bibliography{bibi}

\end{document}